\documentclass[sigconf]{acmart}

\def\BibTeX{{\rm B\kern-.05em{\sc i\kern-.025em b}\kern-.08emT\kern-.1667em\lower.7ex\hbox{E}\kern-.125emX}}

\usepackage{xspace}

\newcommand{\pMA}{MAP\xspace}
\copyrightyear{2019}
\acmYear{2019}
\setcopyright{rightsretained}
\acmConference[CF '19]{Proceedings of the 16th conference on Computing Frontiers}{April 30-May 2, 2019}{Alghero, Italy}
\acmBooktitle{Proceedings of the 16th conference on Computing Frontiers (CF '19), April 30-May 2, 2019, Alghero, Italy}
\acmDOI{10.1145/3310273.3323421}
\acmISBN{978-1-4503-6685-4/19/05}

\begin{document}
	
	\title{IMD Security vs. Energy: Are we tilting at windmills?: POSTER}
	\titlenote{This work has been supported by the EU-funded project SDK4ED (Grant Agreement No. 780572).}
	%\subtitle{Extended Abstract}
	%\subtitlenote{The full version of the author's guide is available as
	%  \texttt{acmart.pdf} document}
	
	\author{Muhammad Ali Siddiqi}
	\orcid{0000-0002-8554-7077}
	\affiliation{%
		\institution{Department of Neuroscience, Erasmus Medical Center}
		\city{Rotterdam}
		\state{The Netherlands}
	}
	\email{m.siddiqi@erasmusmc.nl}
	\author{Christos Strydis}
	%\orcid{1234-5678-9012}
	\affiliation{%
		\institution{Department of Neuroscience, Erasmus Medical Center}
		\city{Rotterdam}
		\state{The Netherlands}
	}
	\email{c.strydis@erasmusmc.nl}

	\begin{abstract}
		Implantable Medical Devices (IMDs) such as pacemakers and neurostimulators are highly constrained in terms of energy. In addition, the wireless-communication facilities of these devices also impose security requirements considering their life-critical nature.
		However, security solutions that provide considerable coverage are generally considered to be too taxing on an IMD battery.
		Consequently, there has been a tendency to adopt ultra-lightweight security primitives for IMDs in literature.
		In this work, we demonstrate that the recent advances in embedded computing in fact enable the IMDs to use more mainstream security primitives, which do not need to compromise significantly on security for fear of impacting IMD autonomy.
	\end{abstract}
	
	%
	% The code below is generated by the tool at http://dl.acm.org/ccs.cfm.
	% Please copy and paste the code instead of the example below.
	%
	
	
	%
	% Keywords. The author(s) should pick words that accurately describe the work being
	% presented. Separate the keywords with commas.
	
	\maketitle
	
	\section{Introduction}
\label{sec:introduction}

Many battery-powered implantable medical devices (IMDs) are designed to operate for around ten years while implanted in the body.
In order to prolong battery life, traditionally there has been a tendency to opt for less demanding and simple medical-application implementations and underlying hardware.
This is especially important given the safety-critical nature of these devices and their extreme reliability and availability demands. 
Unfortunately, until very recently, the above tendency resulted in less focus on the security aspects of IMDs given the notorious nature of the security solutions in terms of additional energy costs.
Security, as a requirement, has been systematically ignored -- or at the very least, postponed -- despite the fact that IMDs have been increasingly getting equipped with wireless connectivity to the outside world.

Over the last decade or so, numerous vulnerabilities have been found in IMDs due to their lack of security~\cite{halperin2008pacemakers,marin2018security}.
Consequently, there has been a significant ramp up in the number of solutions proposed in literature to strengthen IMD security. However, due to the stringent energy constraints of IMDs, most of the focus has been on developing \emph{lightweight} security solutions~\cite{strydis2013system,marin2018security}. This energy efficiency, however, comes at the cost of not adopting more mainstream and well-scrutinized solutions. One significant downside of this trend is that the proposed solutions do not provide the fundamental \emph{security services}, which one would expect to be provided in life-critical systems such as IMDs.
What is more, adopting well-established solutions would benefit re-certification of now more secure implants through wide-spread security techniques.

From the CIANA\footnote{CIANA: Confidentiality, Integrity, Authentication, Non-repudiation and Availability} security-service model, \emph{confidentiality}, \emph{integrity} and \emph{authentication} between the IMD and an external reader/\-programmer are addressed through the use of lightweight block ciphers in most of the recent work on IMD security~\cite{strydis2013system}.
However, these symmetric primitives, by definition, do not provide \emph{true} non-repudiation, and require a pre-shared key between the reader and IMD.
Asymmetric (or public-key) cryptography primitives, which do not have such issues, have been generally avoided in the case of IMDs for two reasons: They require the use of certificates and a public key infrastructure (PKI) in order to protect against man-in-the-middle (MITM) attacks \emph{and} they are generally considered to be too taxing on the battery~\cite{strydis2013system}. Regarding the first reason, IMDs do not have sufficient on-board memory and lack an Internet connection to track the validity of the reader certificates~\cite{marin2018security}.
This, however, can be addressed by employing a policy that establishes proximity between the IMD and the reader in order to get rid of the need for certificates. This principle, which is called \emph{touch-to-access}, ensures that only the entity that can physically touch the patient is allowed access to the IMD~\cite{rostami2013heart}. The underlying assumption is that the attacker cannot get in close proximity to the IMD since the patient would reject prolonged physical contact with strangers.

\begin{figure}
	\centering
	\includegraphics[trim={0.4cm 4.2cm 0.8cm 0.2cm},clip,width=0.4\textwidth]{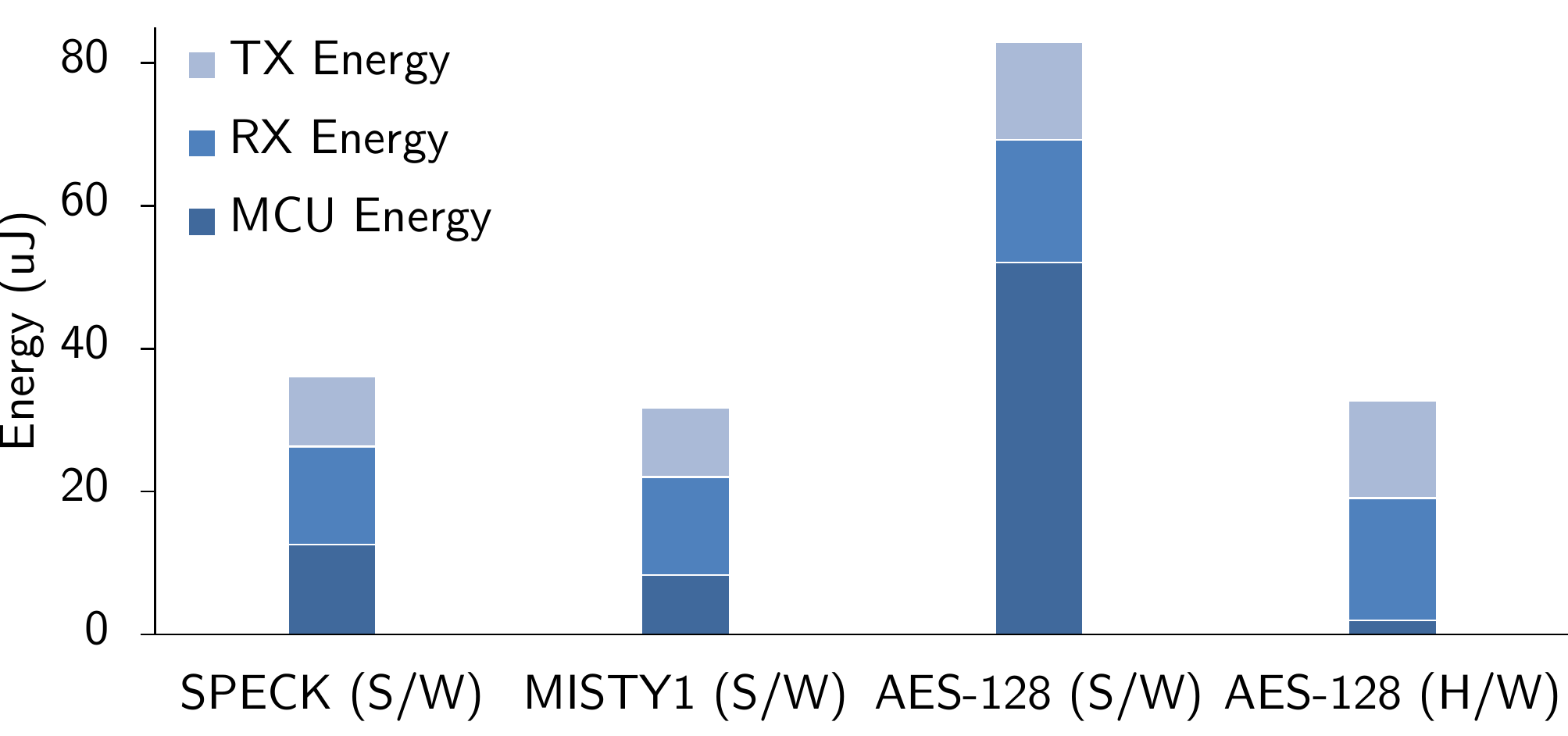}
	\vspace{-0.3cm}
	\caption{Cost of one \pMA session with respect to the choice of cryptographic primitive}
	\label{fig:cipher-comparison}
	\vspace{-0.2cm}
\end{figure}

\begin{figure}
	\centering
	\includegraphics[trim={0.5cm 3.4cm 0.7cm 0.2cm},clip,width=0.4\textwidth]{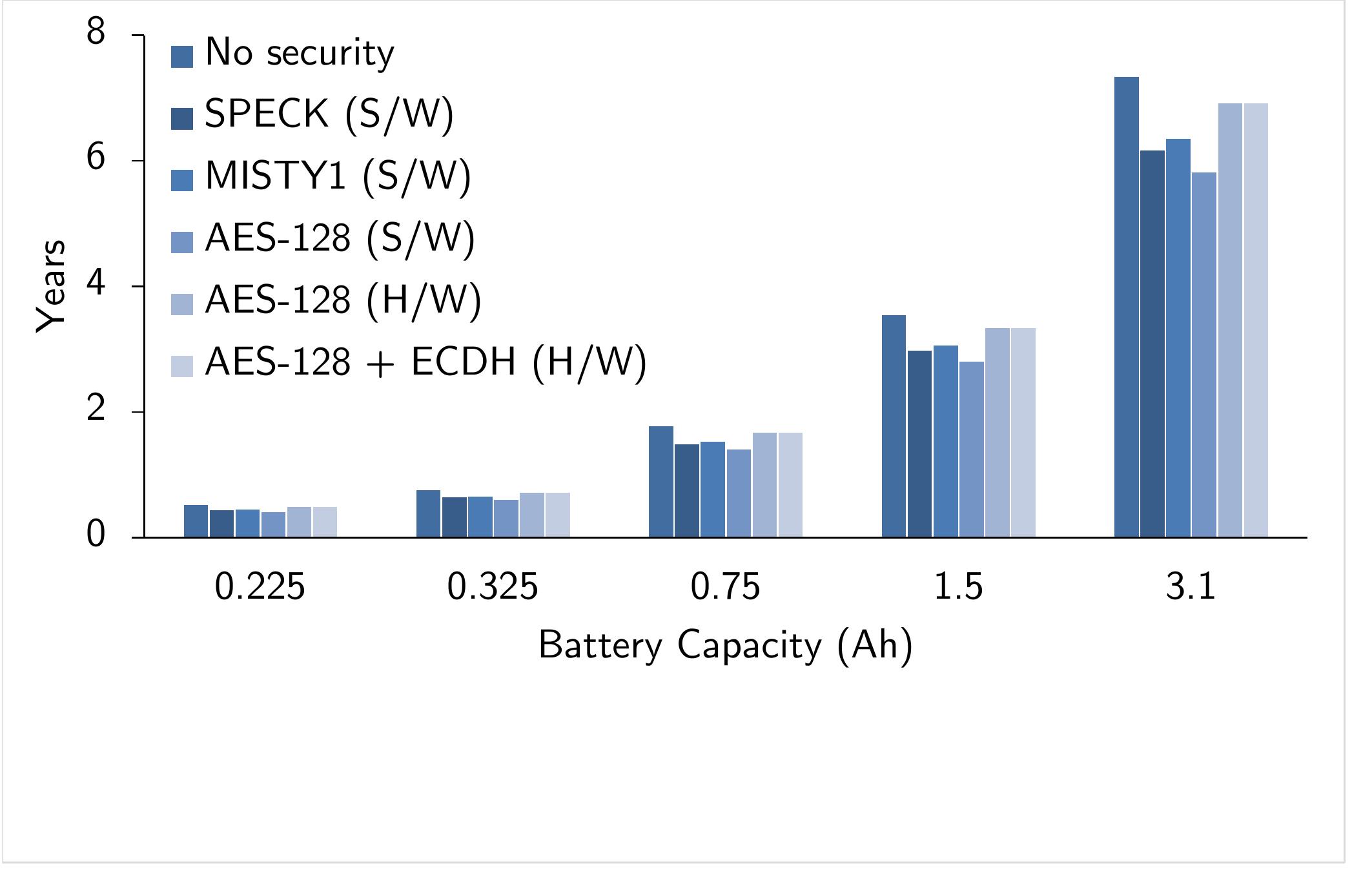}
	\vspace{-0.3cm}
	\caption{IMD-battery lifetime with respect to the choice of cryptographic primitive}
	\label{fig:battery-lifetime}
	\vspace{-0.2cm}
\end{figure}

In order to address the second reason, we evaluate in this article the effect of employing both symmetric (lightweight and non-lightweight) and asymmetric cryptographic primitives on the IMD battery life.
We perform this evaluation on a modern ultra-low-power ARM Cortex-M0+ based MCU~\cite{tinygecko}, which is certainly eligible for use in IMDs.
This MCU has a \emph{dedicated peripheral} for cryptographic acceleration, which is used here on purpose for adding a comparison of hardware-based implementations in the measurements.
As an example, we use an ISO/IEC 9798-2-based, three-pass mutual authentication protocol (\pMA) between the reader and IMD~\cite{strydis2013system}. The protocol employs a symmetric cipher for confidentiality, integrity and authentication and assumes a shared key between the two entities. 
Similarly to~\cite{strydis2013system}, we use separate MCUs for executing the primary medical functionality and security tasks, respectively. This dual-processor architecture helps in handling repeated communication requests, which otherwise may block the implant from performing its primary functionality.
Figure~\ref{fig:cipher-comparison} shows the costs of one \pMA session when using different cipher implementations for the MCU clock frequency of 19 $MHz$.
One \pMA session involves the reader sending a 32-bit command and the implant replying with a 64-bit response.
In this evaluation we executed C implementations of two lightweight block ciphers SPECK and MISTY1, and a non-lightweight -- yet widely-used -- AES-128 on the MCU's CPU.
We also executed the protocol by running AES-128 on the dedicated peripheral.
The wireless-communication energy numbers are based on an implantable-grade transceiver~\cite{microsemi70103}, with an assumed effective data rate of 265 $kbps$.
We see that the software implementation of AES-128 consumes more energy compared to the lightweight ciphers. However, the dedicated hardware implementation of AES-128 consumes the lowest energy across the board.
To demonstrate the impact of public-key cryptography, we use Elliptic-curve Diffie-Hellman (ECDH) for establishing the 128-bit shared symmetric key for \pMA.
For instance, when using the secp256r1 curve, executing ECDH key exchange on the dedicated peripheral before starting the MAP protocol adds 1457 $\mu J$ to the security energy budget needed. Undoubtedly, this is a significant increase, and somewhat justifies the above-mentioned hesitance of not employing public-key cryptography in IMDs.
However, to holistically study the impact on the IMD lifetime, the usage pattern of the IMDs and the battery capacity should be taken into account.
Figure~\ref{fig:battery-lifetime} shows the impact of different \pMA implementations on the battery lifetime compared to the case of having no security at all, for different implantable-grade battery sizes.
The medical-functionality-MCU duty cycle is set to 5\% (active vs. sleep mode), which is in line with a typical pacemaker design~\cite{lindqvist2005compression}.
The communication-session duration is set as 2 minutes per day to match an actual reader-IMD session~\cite{merlin2015faq}.
However, the active-transmission is pessimistically enabled \emph{throughout} the session for worst-case numbers.
We observe that \pMA based on the hardware implementation of AES-128 performs best in terms of IMD longevity. Moreover, the additional cost of ECDH for key exchange is insignificant in the context of IMD lifetime since it is required only once before each daily transmission.

\begin{figure}
	\centering
	\includegraphics[trim={0.4cm 2.8cm 0.7cm 0.2cm},clip,width=0.4\textwidth]{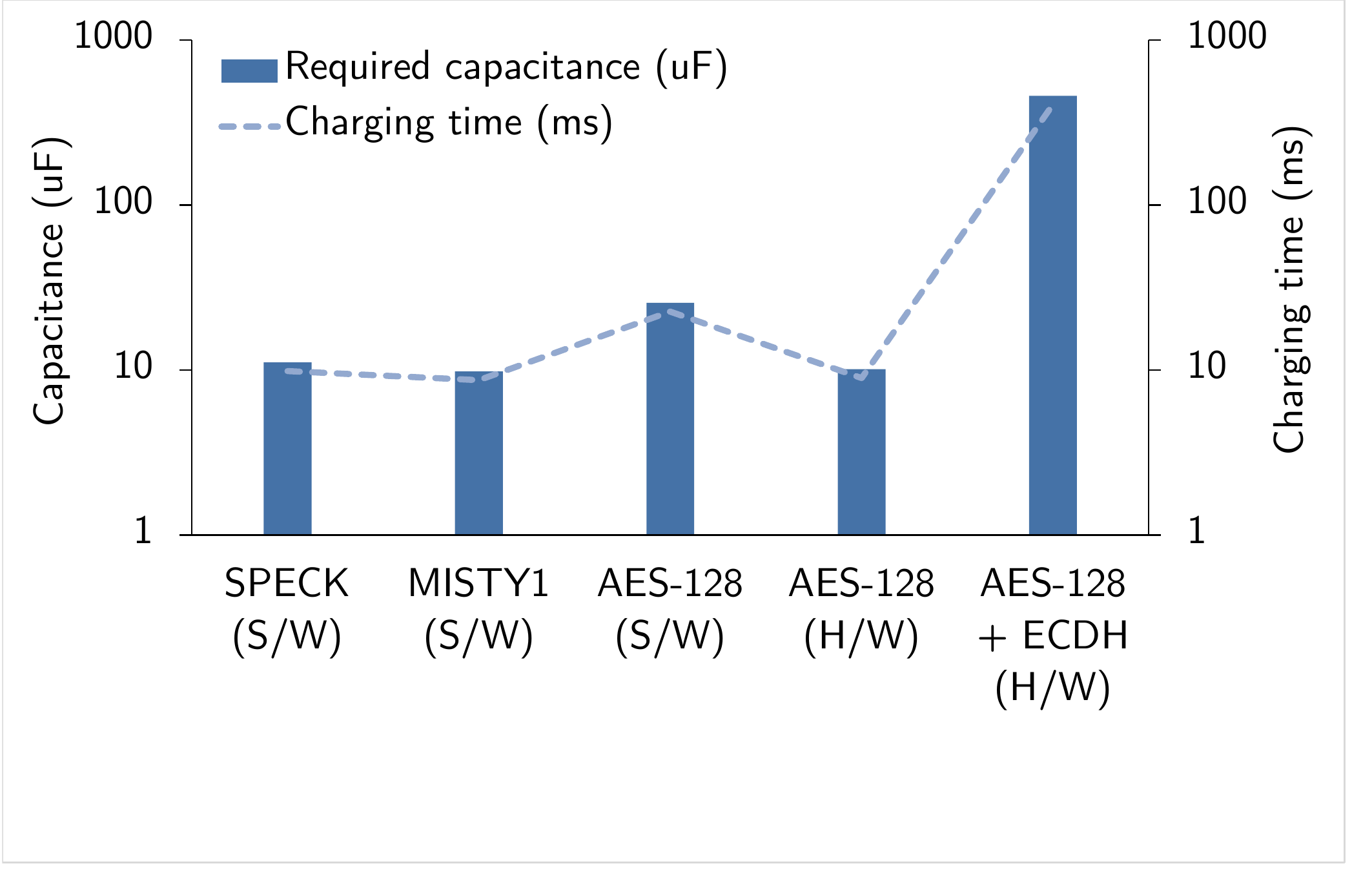}
	\vspace{-0.3cm}
	\caption{Required ZPD capacitance and charging time}
	\label{fig:capacitance-charging-time}
	\vspace{-0.2cm}
\end{figure}

When it comes to \emph{availability}, protection against battery Denial-of-Service (bDoS) attacks is of extreme significance given their low cost and high likelihood in the context of IMDs.
In bDoS, the attacker forces the IMD to run an energy-consuming task (such as authentication), by e.g., sending continuous communication requests, with the aim of depleting its battery.
One effective technique against this attack is \emph{zero-power defense} (ZPD)~\cite{halperin2008pacemakers}, in which the IMD harvests the RF energy from the initiator's communication messages and utilizes this \emph{free} energy to perform authentication.
Only after successful authentication does the IMD switch to battery power.
The simplest configuration of ZPD employs a capacitor that first stores the required energy and then performs authentication~\cite{cypress2017}.
The use of non-lightweight encryption primitives discussed above may require a large capacitance size due their high energy consumption. This can potentially increase the IMD size and also impact real-time performance due to the charging delay.
However, we find that the required capacitance size for AES-128-based \pMA along with ECDH results in roughly 460 $\mu F$ (see Figure~\ref{fig:capacitance-charging-time}), which is within the range of commercial ceramic capacitors\footnote{Ceramic capacitors have a low leakage current, small size and low cost and hence are ideal for energy harvesting~\cite{cypress2017}.}.
The minimum and maximum capacitor voltages are set to 2.1 $V$ and 3.3 $V$, respectively.
By choosing a standard available size of 470 $\mu F$ and using the wireless-power-transfer scheme from~\cite{li2007wireless} as an example (which delivers 6.15 $mW$), the charging delay results in 416 ms, which does not have a practical impact in IMD functionality or availability.

\section{Result and Conclusion}
\label{sec:conclusion}

Based on the findings in this work, we conclude that compromising security-service coverage by choosing a lightweight solution is not a recommended move considering the current state of the art of ultra-low-power embedded computing.
Our recommendation is to focus more on extending this coverage and less on the impact on battery life, since our results show that modern MCUs allow the use of mainstream primitives in an energy-efficient manner.
	
%	\begin{acks}
%		This work has been supported by the EU-funded project SDK4ED (Grant Agreement No. 780572).
%	\end{acks}
	
	\bibliographystyle{ACM-Reference-Format}
	\bibliography{ms}
	
\end{document}